\documentclass[journal]{IEEEtran}
\usepackage{cite}
\usepackage{array}
\usepackage{longtable}
\usepackage{multirow}
\usepackage{amsmath}
\usepackage{geometry}
\usepackage{graphicx}
\usepackage{tikz}
\usepackage{enumitem}
\usepackage{booktabs}
\usetikzlibrary{shapes.geometric, arrows, positioning}

\tikzstyle{cause} = [rectangle, rounded corners, draw, align=center, text width=5cm, minimum height=2cm]
\tikzstyle{topcause} = [rectangle, draw, align=center, text width=5cm, minimum height=2cm]
\tikzstyle{arrow} = [thick,->,>=stealth]
\newgeometry{top=20mm, bottom=20mm, left=17mm, right=17mm}

\ifCLASSINFOpdf
\else
\fi
\usepackage{algorithmic}

\usepackage{array}

\begin{document}
%
\title{The Necessity of a Holistic Safety Evaluation Framework for AI-Based Automation Features}
%
%

\author{Alireza~Abbaspour $^{*,a}$,~
        Shabin~ Mahadevan $^{a}$,~
        ~Kilian~ Zwirglmaier $^{a}$,~and Jeff Stafford $^{a}$
\thanks{$^{a}$ Qualcomm Technologies Inc., San Diego, CA 92121 USA.\\
$^{*}$ Corresponding author: aabbaspo@qti.qualcomm.com (A.Abbaspour)}
}

%
%

\markboth{Pre-Print for SAE WCX 2026 Conference\ }%
{Shell \MakeLowercase{\textit{et al.}}: Bare Demo of IEEEtran.cls for IEEE Journals}
%



\maketitle

\begin{abstract}
The intersection of Safety of Intended Functionality (SOTIF) and Functional Safety (FuSa) analysis of driving automation features has traditionally excluded Quality Management (QM) components (components that has no ASIL requirements allocated from vehicle-level HARA) from rigorous safety impact evaluations. While QM components are not typically classified as safety-relevant, recent developments in artificial intelligence (AI) integration reveal that such components can contribute to SOTIF-related hazardous risks. Compliance with emerging AI safety standards, such as ISO/PAS 8800, necessitates re-evaluating safety considerations for these components. This paper examines the necessity of conducting holistic safety analysis and risk assessment on AI components, emphasizing their potential to introduce hazards with the capacity to violate risk acceptance criteria when deployed in safety-critical driving systems, particularly in perception algorithms. Using case studies, we demonstrate how deficiencies in AI-driven perception systems can emerge even in QM-classified components, leading to unintended functional behaviors with critical safety implications. By bridging theoretical analysis with practical examples, this paper argues for the adoption of comprehensive FuSa, SOTIF, and AI standards-driven methodologies to identify and mitigate risks in AI components. The findings demonstrate the importance of revising existing safety frameworks to address the evolving challenges posed by AI, ensuring comprehensive safety assurance across all component classifications spanning multiple safety standards. 
\end{abstract}


%
\IEEEpeerreviewmaketitle

\section{Introduction}
Ensuring safety is the most critical topic in  Advanced Driver Assistance Systems (ADAS) as these systems progressively control critical driving tasks. ADAS technologies, including hands-off driving, lane-keeping assistance, adaptive cruise control, and automated emergency braking (AEB), operate in dynamic and unpredictable environments, where even minor safety deficiencies can lead to catastrophic outcomes. As these systems evolve toward higher levels of autonomy, the complexity of their architectures increases drastically. This complexity is further magnified by the ongoing rapid adoption of artificial intelligence (AI) elements, particularly in perception and decision-making modules.\\

The application of AI in ADAS introduces both opportunities and challenges. While AI-driven models enhance capabilities such as object detection, semantic segmentation, and prediction of pedestrian intent, they also present new safety concerns. These concerns stem from the inherent opaqueness and stochastic nature of AI algorithms, which can result in unpredictable or unintended system behavior \cite{pitale2024inherent}. Traditional Functional Safety (FuSa) analysis, as defined by ISO 26262 \cite{ISO26262}, focuses on systematic and random hardware and software failures. However, it is not designed to address the performance limitations and functional insufficiencies characteristic of AI-based components. To bridge this gap, the Safety of Intended Functionality (SOTIF), governed by ISO 21448 \cite{ISO21448}, has emerged as a critical framework. SOTIF extends beyond traditional FuSa by addressing hazards arising from functional insufficiencies and foreseeable misuse.\\

Despite the utility of SOTIF, there is a persistent misunderstanding that components classified under QM based on FuSa analysis are inherently free from SOTIF‑related hazards. QM‑rated AI components are often excluded from rigorous safety evaluations due to their classification outside the scope of safety‑critical analysis; however, this assumption is increasingly untenable as AI‑based components, even those designated as QM‑rated, demonstrate the potential to introduce functional insufficiencies, particularly in AI‑based perception systems, where subtle inaccuracies in model predictions or training data biases can lead to hazardous outcomes in real‑world scenarios. System‑Theoretic Process Analysis (STPA) is a systems‑theoretic hazard analysis approach that addresses both Functional Safety (FuSa) and Safety of the Intended Functionality (SOTIF) concerns within complex socio‑technical control structures. Developed by Nancy G. Leveson at MIT \cite{leveson2018stpa}, STPA systematically models controllers, actuators, sensors, human operators, and the operating environment to identify potential unsafe control actions (UCAs), derive enforceable safety constraints, and generate causal scenarios; as such, it has been widely adopted in the aerospace and automotive domains for comprehensive identification of both SOTIF‑ and FuSa‑related hazards \cite{haixia2022complete,chen2025stpa}.
 
While the emergence of the ISO/PAS 8800 standard for AI safety \cite{ISO8800} brings much-needed clarity to assurance practices for AI-driven systems, it also introduces new dimensions to an already complex autonomous driving landscape.  This paper explores the reasoning behind the necessity of conducting SOTIF analysis on QM rated AI components. By examining use cases from AI-based perception systems, it highlights the critical need for a revised approach to safety assurance when it comes to QM rated AI components. This analysis is vital to ensure that the growing complexity and autonomy of ADAS systems are matched by robust methodologies capable of identifying and mitigating emerging risks.
This paper makes the following key contributions to the field of safety analysis in ADAS systems:\\
1.	It demonstrates the theoretical necessity of evaluating QM rated AI components for safety-related hazards.
2.	It provides practical use cases illustrating how QM rated AI components can contribute to SOTIF-related hazards.
3.	It argues and exemplifies why AI safety must address SOTIF aspects of QM rated AI components.
By addressing these areas, this work advocates for a more inclusive and robust safety framework that aligns with the growing complexity and autonomy of modern ADAS systems.\\

The rest of this paper is outlined as follows: Section II examines the interaction between FuSa and SOTIF, emphasizing the role of ASIL ratings and demonstrating that QM-rated AI components can contribute to SOTIF-related hazards. Section III presents an AI-based perception use case in a Level 2+ ADAS system, showcasing how QM-rated AI components can lead to safety risks. Section IV  demonstrates how to ensure AI safety using ISO/PAS 8800 guidelines in a safety critical AI-based perception system. Section V discusses the findings, reinforcing the need for holistic safety analysis of AI-based systems. Section VI concludes the paper and outlines future research directions.

\section{FuSa and SOTIF Interaction at ASIL}
The interaction between FuSa and SOTIF is pivotal for ensuring comprehensive safety in ADAS systems. FuSa, defined by ISO 26262, addresses failures arising from systematic and random faults, while SOTIF, governed by ISO 21448, focuses on hazards resulting from functional insufficiencies or foreseeable misuse. Together, these frameworks provide a holistic approach to hazard identification and mitigation. The goal of this section is to discuss that, despite the ASIL rate allocated to a component or sub-system, a SOTIF hazard may still occur even if that component or sub-system is considered as QM.
\subsection{FuSa Aspects of Safety}
 Automotive Safety Integrity Levels (ASIL) are a key component of FuSa, categorizing risks based on severity (S), exposure (E), and controllability (C). ASIL levels range from A (lowest) to D (highest), with QM rating representing non-critical components outside the ASIL framework. SOTIF analysis often intersects with these ratings, particularly in identifying hazards that arise despite the absence of E/E failures. Conditions for SOTIF hazards are described by controllability and severity, emphasizing scenarios where hazards occur with $C > 0$ and $S > 0$.
This section examines how components classified as QM  in FuSa can still contribute to SOTIF-related hazards. For instance, a perception module designated as QM may fail to correctly identify road objects under specific conditions, leading to a hazardous scenario despite no hardware or software faults. These examples illustrate the need to evaluate QM rated AI components under the SOTIF framework to ensure that potential hazards are identified and mitigated effectively.
FuSa in the automotive industry, as outlined in the ISO26262 standard, refers to minimizing unreasonable risks caused by the malfunction of electrical or electronic systems. This standard seeks to prevent injuries or harm to people’s health by addressing risks in vehicle components. In the automotive industry, according to ISO 26262 \cite{ISO26262}, risk is divided into three components: Severity, Occurrence (referred to as “Exposure”), and Controllability. Each of these components is further categorized into various qualitative levels, as detailed in the following list:
\begin{itemize}
\item Severity (S)\\
- $S0$: No injuries\\
- $S1$: Light and moderate injuries\\
- $S2$: Severe and life-threatening injuries (survival probable)\\
- $S3$: Life-threatening injuries (survival uncertain)
\item Exposure (E)\\
- $E1$: Very low probability\\
- $E2$: Low probability\\
- $E3$: Medium probability\\
- $E4$: High probability
\item Controllability (C)\\
- $C0$: the hazardous event is controllable in general\\
- $C1$: Simply controllable\\
- $C2$: Normally controllable\\
- $C3$: Difficult to control or uncontrollable
\end{itemize}
By combining qualitative estimates for Severity, Exposure, and Controllability, a risk level is determined, which corresponds to an ASIL. Only the highest risk combinations qualify for ASIL D, while most low-risk combinations do not qualify for any ASIL, which are referred to QM.
Some examples to better understand this  concept are, ASIL D (the highest level) includes risks such as unintended airbag deployment and certain types of unwanted deceleration or self-steering failures. ASIL C covers some types of unintended  braking, or acceleration. ASIL B includes failures in front or rear-view cameras or brake lights. ASIL A involves failures of rear lights on both sides. As automation increases, human drivers will be less able to assist with controllability, potentially moving some systems to higher ASILs. Table I shows how Severity, Controllability and Exposure are used to determine the ASIL rating:\\
\begin{table}[h!]
\caption{ASIL determination table based on severity, probability, and controllability classes.}
\scriptsize
\centering
\renewcommand{\arraystretch}{1.4}
\setlength{\tabcolsep}{8pt}
\begin{tabular}{|c|c|c|c|c|}
\hline
\multirow{2}{*}{\textbf{Severity Class}} & \multirow{2}{*}{\textbf{Probability Class}} & \multicolumn{3}{c|}{\textbf{Controllability Class}} \\ \cline{3-5} 
                                         &                                             & \textbf{C1} & \textbf{C2} & \textbf{C3} \\ \hline
\multirow{4}{*}{S1}                      & E1                                          & QM           & QM           & QM           \\ \cline{2-5} 
                                         & E2                                          & QM          & QM          & QM          \\ \cline{2-5} 
                                         & E3                                          & QM          & QM          & A           \\ \cline{2-5} 
                                         & E4                                          & QM          & A           & B           \\ \hline
\multirow{4}{*}{S2}                      & E1                                          & QM          & QM          & QM          \\ \cline{2-5} 
                                         & E2                                          & QM          & QM          & A           \\ \cline{2-5} 
                                         & E3                                          & QM          & A           & B           \\ \cline{2-5} 
                                         & E4                                          & A           & B           & C           \\ \hline
\multirow{4}{*}{S3}                      & E1                                          & QM          & QM          & A           \\ \cline{2-5} 
                                         & E2                                          & QM          & A           & B           \\ \cline{2-5} 
                                         & E3                                          & A           & B           & C           \\ \cline{2-5} 
                                         & E4                                          & B           & C           & D           \\ \hline
\end{tabular}
\label{table:ASIL}
\end{table}

\subsection{SOTIF Aspects of Safety: }
SOTIF is a critical concept in the realm of automated driving systems, aimed at ensuring these systems operate safely under all intended conditions. Unlike traditional FuSa, which primarily addresses risks stemming from hardware and software malfunctions, SOTIF focuses on hazards arising from functional insufficiencies, foreseeable misuse by drivers, performance limitation, and unexpected conditions in combination with triggering conditions within the operational design domain (ODD). For example, SOTIF would address scenarios such as a sensor's inability to detect an obstacle under specific lighting conditions. The ISO 21448 standard provides comprehensive guidelines for achieving SOTIF, emphasizing the importance of hazard analysis, risk assessment, and the implementation of safety measures throughout the system's lifecycle. By integrating SOTIF with FuSa, manufacturers can significantly enhance the overall safety and reliability of automated vehicles.\\
According to the ISO 21448 standard, if the Controllability (C) is greater than 0 or the Severity (S) of potential harm is greater than 0, it constitutes a residual risk. This necessitates further investigation to specify acceptance criteria for residual risks, as illustrated in Figure 1.
As illustrated in Table \ref{table:ASIL}, for FuSa, as long as the controllability  and severity are both less than or equal to 1, the ASIL rating will be classified as QM, regardless of the exposure level. Consequently, if there are any SOTIF hazards within this region ($0<C\leq 1$ or $0 < S\leq 1$), there will be no corresponding FuSa hazard.

\begin{figure}[h]
    \centering
    \includegraphics[width=0.45\textwidth]{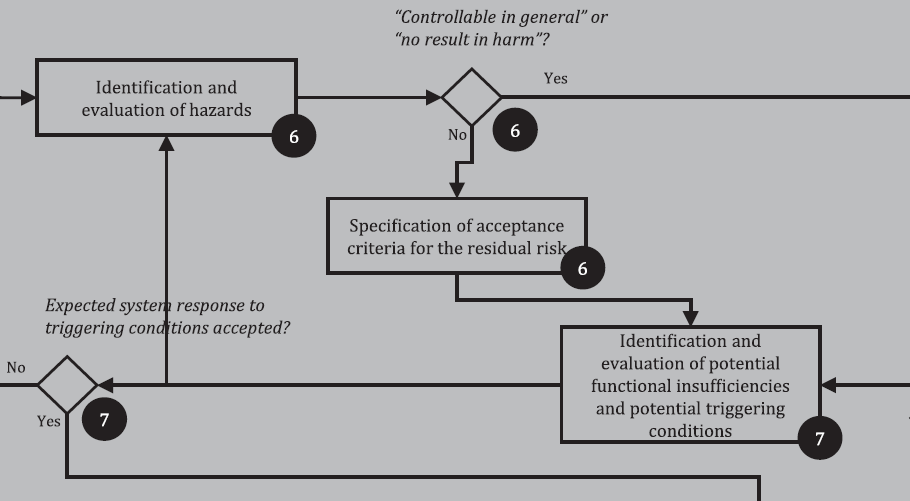}
    \caption{SOTIF Risk Evaluation according to ISO 21448 standard \cite{ISO21448}}
    \label{fig:SOTIF_Risk}
\end{figure}

\subsection{ Identification of SOTIF-Related Hazards in QM rated AI Components }

Some individuals mistakenly believe that QM-rated components are not safety-critical and thus irrelevant to SOTIF. However, all QM components, especially those involving AI in autonomous driving must be carefully assessed for SOTIF relevance. AI-based QM components, in particular, require additional scrutiny and a comprehensive SOTIF analysis to ensure safety risks are properly addressed, even if their FuSa classification remains unchanged.\\

In the following sections, we will present a use-case of a QM rated AI rating that have associated SOTIF hazards. These examples are provided to clarify the intention of this paper, demonstrating how SOTIF hazards can impact components traditionally considered to have negligible safety risks under FuSa definitions. By examining these cases, we aim to highlight the importance of re-evaluating QM rated AI components to ensure comprehensive safety measures are implemented, addressing both FuSa and SOTIF hazards effectively.

\section{QM rated AI Components with SOTIF Hazard Impacts}

In this section, we will use the Low-Level Perception (LLP) system as a case study to demonstrate how  QM rated AI functions can be associated with SOTIF hazards. Specifically, we will examine LLP within a Level 2 system, which is classified as a QM rated AI component. AI-based algorithms are common in modern LLP designs, necessitating additional standard compliance, such as ISO/PAS 8800. By exploring this example, we aim to illustrate the potential SOTIF hazards associated with QM-rated AI components and clarify the importance of addressing these hazards to ensure comprehensive safety.\\

\begin{figure}[t!]
    \centering
    \includegraphics[width=0.48\textwidth]{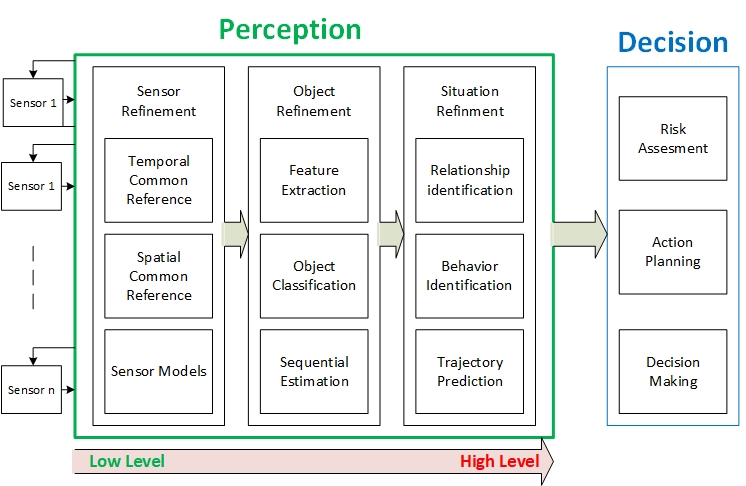}
    \caption{Graphical Description of Low and High level perception role in ADAS application}
    \label{fig:Perception}
\end{figure}

LLP and High-Level Perception (HLP) differ primarily in their scope and complexity. LLP involves the initial processing of raw sensory data, such as detecting edges, textures, and basic shapes from visual inputs. It focuses on fundamental features and patterns that form the building blocks for more complex interpretations. In contrast, HLP integrates and interprets these basic features to recognize objects, understand scenes, and make higher-order decisions. HLP leverages the processed data from LLP to form a comprehensive understanding of the environment, enabling more sophisticated tasks like object classification, scene analysis, and contextual reasoning. Essentially, LLP provides the foundational data, while HLP builds upon it to achieve a deeper and more meaningful understanding of the surroundings. Accurate detection and description of these features are crucial for various computer vision tasks, including object recognition and scene understanding \cite{freeman2000learning}. Figure \ref{fig:Perception} shows a graphical description of perception in ADAS and the roles of LLP and HLP within it.

\subsection{Role of LLP in ADAS}
LLP plays a crucial role in Advanced Driver Assistance Systems (ADAS) by providing the foundational data necessary for various safety and automation functions. LLP processes raw sensory inputs from cameras, radar, and lidar to detect and identify objects such as vehicles, pedestrians, and road signs. This initial object detection is vital for real-time decision-making and collision avoidance. Additionally, LLP contributes to determining the Operational Design Domain (ODD) by continuously monitoring and interpreting environmental conditions to ensure the vehicle operates within its safe limits \cite{malligere2023object}. Furthermore, LLP aids in predicting the ego vehicle's path by analyzing the vehicle's current position, speed, and trajectory, which is essential for Adaptive Cruise Control (ACC), Lane-Keeping Assistance (LKA), Automated Emergency Braking (AEB), and other automated driving tasks \cite{murthy2022objectdetect}. \\

Importantly, LLP often relies on components rated as QM under functional safety standards. While these components are traditionally considered non-safety-critical, their role in perception tasks, especially in AI-driven systems, demands careful evaluation. Machine learning models, particularly convolutional neural networks (CNNs), further enhance LLP by learning to detect these features accurately \cite{freeman2000learning, chen2023can, fan2020deeper, melotti2023reducing}.

\subsection{Role of Machine Learning in LLP} 
Machine learning, particularly through CNNs, has significantly advanced the field of LLP.  CNNs utilize convolutional layers to apply filters that automatically learn to detect and extract low-level features from images. This process enhances the accuracy and efficiency of feature detection, which is essential for tasks such as object recognition and image segmentation \cite{chen2023can}. 

Many of the hardware and software components supporting machine learning in LLP are QM-rated. These include image preprocessing units, embedded AI accelerators, and data pipelines that feed into perception models. Although classified as QM, their contribution to safety-relevant perception tasks in ADAS—especially in autonomous driving—requires special attention. These components must undergo thorough SOTIF evaluation to ensure that their behavior under edge cases or novel scenarios does not compromise system safety. Moreover, machine learning models can adapt to different types of images and environments, making them versatile across various applications, from medical imaging to autonomous driving. By training on large datasets, these models improve their performance and robustness, enabling real-time applications like environmental perception \cite{fan2020deeper}.

\subsection{SOTIF Hazards in LLP for ADAS systems}
To assess whether such hazards are tolerable, ISO 21448 prescribes acceptance rationales based on either qualitative or quantitative thresholds. These are grounded in principles such as GAMAB (“Globalement Au Moins Aussi Bon” / “Globally At Least As Good”), PRB (“Positive Risk Balance”), ALARP (“As Low As Reasonably Practicable”), and MEM (“Minimal Endogenous Mortality”). These principles ensure that the residual risk of the automated function is no greater than that of a competent human driver (or incumbent system), and that any newly introduced risks are outweighed by the net safety benefit. These global acceptance criteria are then decomposed into scenario-class-specific validation targets—concrete, measurable performance objectives for virtual simulation, track, or on-road testing, providing the statistical confidence needed to demonstrate compliance within the defined Operational Design Domain \cite{hofmeister2023acceptance}. By systematically mapping each SOTIF hazard, including QM rated, to tailored risk acceptance criteria and their corresponding validation targets, developers can structure rigorous validation strategies and produce robust evidence that all SOTIF-related risks have been addressed in accordance with ISO 21448’s requirements.\\
Building on the acceptance framework, the decomposition of hazards through CTA (Cause Tree Analysis) yields discrete root‐cause nodes that can be directly traced to specific triggering events and operational situations. By mapping each leaf cause to its corresponding scenario class, teams can define precise, scenario‐based validation targets, such as required coverage levels and statistical confidence bounds, and allocate acceptance thresholds tailored to each hazard pathway. This traceability ensures that mitigation measures not only address underlying system deficiencies but are also verifiable through targeted simulation runs and on‐road tests, thereby closing the loop between hazard identification and the quantitative validation requirements of ISO 21448 \cite{rau2019approach}.

\subsection{Hands off driving}
Hands-off driving (HOD) ADAS features enable the driver to remove their hands from the steering wheel while the system maintains lateral control within a strictly defined Operational Design Domain (ODD), for example, limited to divided highways with clear lane markings and moderate speeds \cite{exposito2024safety}. LLP subsystems leverage multi-sensor fusion and detection algorithms to continuously monitor key ODD parameters, such as lane boundary quality and roadway geometry, and only permit HOD activation when confidence in these conditions is high \cite{abdel2024matched}. However, if LLP falsely assesses the environment as satisfying ODD requirements (a false positive), HOD may engage erroneously, prompting drivers to take their hands off when manual steering intervention remains necessary, an error that can compromise safety and increase collision risk. To mitigate such hazards, developers must adopt conservative gating logic that prioritizes minimizing false positive ODD detections, validated through exhaustive edge-case testing and cross-checking across redundant sensor modalities.\\

\begin{figure*}[h!]
    \centering
    \includegraphics[width=0.9\textwidth]{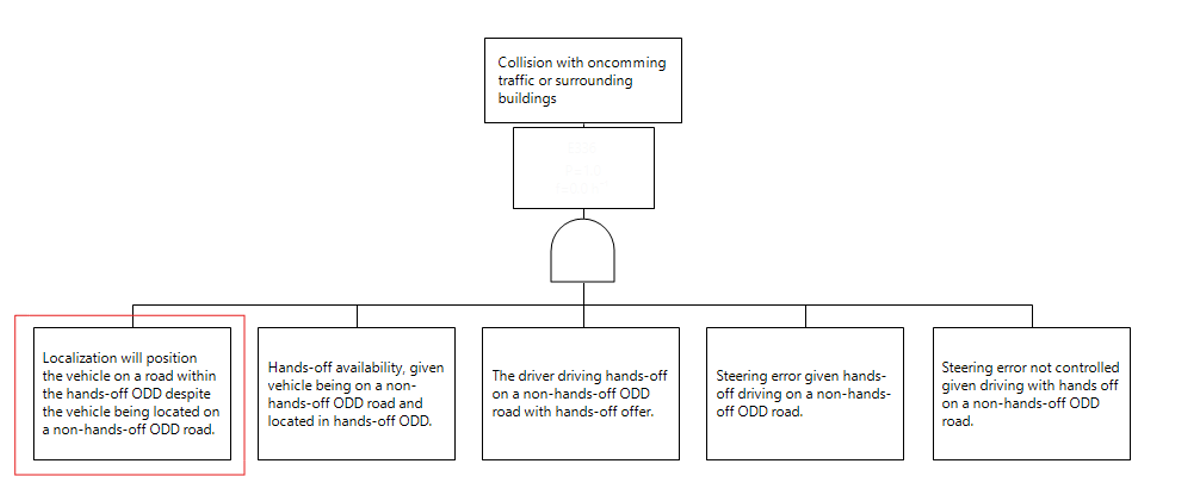}
    \caption{Caust Tree Analysis for Collision with oncoming traffic or surrounding}
    \label{fig:CTA}
\end{figure*}

We conducted a Hazard Analysis and Risk Assessment (HARA) for the HOD feature, focusing specifically on the unintended activation of this function. Given that the driver retains responsibility for vehicle control in a Level 2 system, the associated safety goal was classified as QM, with no safe state required. Table \ref{tab:HARA} summarizes the HARA results.\\
In parallel, a SOTIF Identification and Risk Assessment (SIRA) was performed for the same feature. This analysis identified a SOTIF-related hazard with severity level S2 and controllability level C2, notably consistent with the HARA classification, indicating a potentially unsafe scenario despite the QM rating. Table \ref{tab:sotif_analysis} details the SIRA findings.\\
To further investigate the origin of this hazard, a CTA was conducted (Fig. \ref{fig:CTA}), illustrating how a latent learning problem  could contribute to the identified risk. Because the scenario exhibits S2/C2 in both the HARA and the SIRA, ISO 21448 requires conducting a Residual Risk Assessment (RRA). For the purposes of this discussion, we assume that the preliminary residual‑risk estimate exceeds the prescribed SOTIF acceptance criterion; consequently, additional SOTIF mitigation requirements are necessary to reduce the residual risk below the defined threshold.\\
\begin{table*}[h!]
\centering
\caption{HARA for unintended activation scenario}
\scriptsize
\begin{tabular}{|p{1cm}|p{1cm}|p{1cm}|p{1cm}|c|c|c|p{2cm}|p{1cm}|p{1cm}|}
\hline
\textbf{Action} & \textbf{Hazard} & \textbf{Situation} & \textbf{Hazardous event} & \multicolumn{3}{|c|}{\textbf{Risk Rating}} & \textbf{Safety Goal} & \textbf{ASIL} & \textbf{Safe State} \\ \cline{5-7}
 & & & & \textbf{Severity} & \textbf{Exposure} & \textbf{Controllability} & & & \\ \hline
Hands off activation & Unintended activation & Erroneous activation in roads close to highway & Lateral collision with vehicles & S2 & E2 & C2 & Not safety relevant since driver is responsible for ensuring safe ODD activation & QM & Not applicable \\ \hline
\end{tabular}
\label{tab:HARA}
\end{table*}


\begin{table*}[h!]
\centering
\caption{SOTIF  analysis and Risk Assessment (SIRA) for erroneous ODD detection scenario}
\scriptsize
\begin{tabular}{|p{4.5cm}|p{3cm}|p{2cm}|p{2cm}|p{2cm}|}
\hline
\textbf{Potentially hazardous scenario} & \textbf{Hazardous event} & \multicolumn{2}{c|}{\textbf{Risk rating}} & \textbf{RRA required} \\
\cline{3-4}
& & \textbf{Severity} & \textbf{Controllability} & \\
\hline
Vehicle is driving on a motorway-like road (outside of ODD) that is located close to a motorway (inside of ODD). The vehicle is erroneously located on the motorway and therefore the feature is offered to the driver, who accepts the offer. 
& An incorrectly computed trajectory causes the vehicle to leave the lane and to collide into a vehicle or object. 
& S2 & C2 
& Yes (since $C > 0$ and $S > 0$) \\
\hline
\end{tabular}
\label{tab:sotif_analysis}
\end{table*}
The ADS shall use diverse sensing procedures to ensure the vehicle is geo-fenced within the ODD.
This example illustrates that even a function rated as QM, typically not associated with safety-critical concerns, can still lead to a SOTIF hazard. This challenges the common assumption that QM-classified functions are inherently free from SOTIF-related risks.

\section{ISO/PAS 8800 Safety Aspect of LLP}
\begin{figure}[h!]
    \centering
    \includegraphics[width=0.35\textwidth]{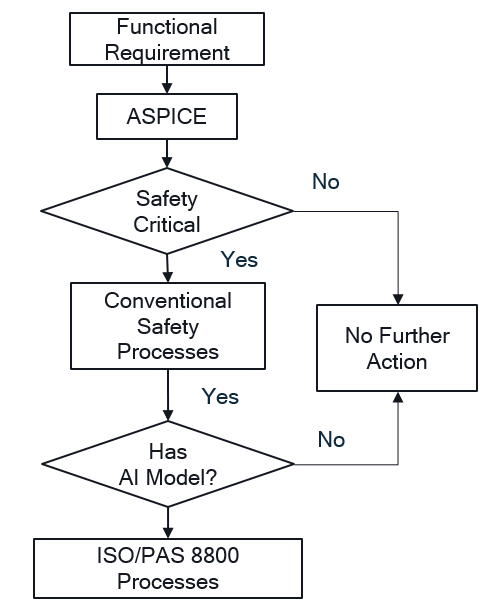}
    \caption{Determining when to apply ISO/PAS 8800.}
    \label{fig:8800_Process}
\end{figure}

In the context of safety-critical systems, the integration of AI introduces unique challenges that necessitate rigorous oversight and standardization. The ISO/PAS 8800 standard provides a structured framework specifically designed to address these challenges. It recommends that its guidelines be followed whenever a safety-critical component incorporates an AI model, particularly when the associated risks can be directly linked to the behavior or performance of the AI system. Importantly, ISO/PAS 8800 is intended as a complementary standard to ISO 26262, which addresses FuSa, and ISO 21448, which focuses on SOTIF. It does not replace these standards but rather extends their applicability to AI-driven components. Figure \ref{fig:8800_Process} graphically illustrates the decision-making process for determining when ISO/PAS 8800 should be applied in automotive applications, emphasizing its relevance when AI contributes to the risk profile of a safety-critical function.

In the following subsections, we aim to summarize the primary aspects of the ISO/PAS 8800 standard and explore how it addresses AI-specific safety concerns, particularly in the context of safety-critical applications LLP in autonomous driving systems. ISO/PAS 8800 provides a structured approach to identifying, analyzing, and mitigating risks introduced by AI components, ensuring that their integration into safety-critical functions does not compromise overall system safety.\\

  \subsubsection{AI Safety Lifecycle}  
\begin{figure}[h!]
    \centering
    \includegraphics[width=0.5\textwidth]{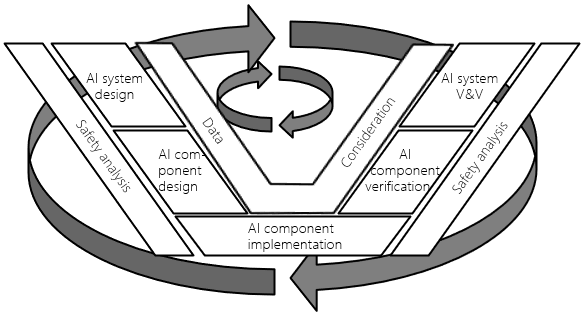}
    \caption{AI system design and V$\&$V phase of the  AI safety lifecycle introduced by ISO/PAS 8800.}
    \label{fig:8800_lifecycle}
\end{figure}

The AI safety lifecycle is an iterative process grounded in the V-model and systems engineering principles, aligned with ISO 8800 phases. It begins with constructing a safety assurance argument and defining AI safety requirements, followed by rigorous data management and model development. Verification and validation ensure the system meets safety expectations both in design and in real-world conditions. Finally, continuous monitoring during operations helps identify and address emerging risks. This lifecycle emphasizes traceability, adaptability, and ongoing improvement to maintain AI system safety throughout its development and deployment. Fig.\ref{fig:8800_lifecycle} shows AI system design and V$\&$V phase of the  AI safety lifecycle introduced by ISO/PAS 8800 \cite{ISO8800}. \\
\begin{figure}[h!]
    \centering
    \includegraphics[width=0.5\textwidth]{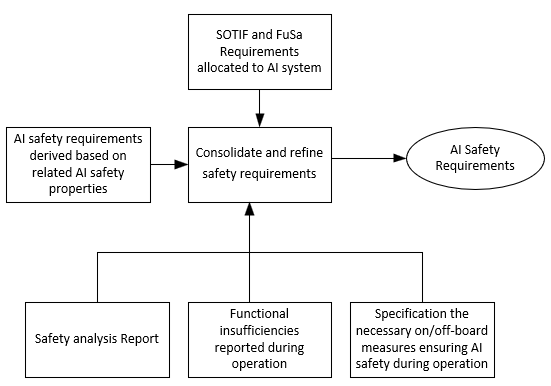}
    \caption{Flow chart for derivation of AI safety requirements.}
    \label{fig:AI_Safety_Req}
\end{figure}
  \subsubsection{Derivation of AI Safety Requirements}  
Derivation of AI safety requirements is one of the first steps toward developing a safe AI system. According to ISO/PAS 8800, it is recommended to derive AI safety requirements by utilizing functional requirements, safety requirements allocated to the AI system, input space definitions, AI safety properties, safety analyses of the AI system, and functional insufficiencies identified during operation \cite{ISO8800}. \

To this end, we present an example illustrating how to derive AI safety requirements from the SIRA analysis shown in Table~\ref{tab:sotif_analysis}. This example demonstrates how SOTIF system requirements can be used in the derivation of AI safety requirements for a geofence (ODD) detection LLP module in an ADS. It highlights how safety properties, hazard analysis, and mitigation strategies can lead to actionable AI safety requirements. While not exhaustive, this example serves as a proof of concept. The following steps illustrate the derivation process, as outlined in Fig.~\ref{fig:AI_Safety_Req}.

\textbf{Step 1. Baseline SOTIF Requirement}\\
The foundational SOTIF requirement is:

“The ADS shall use diverse sensing procedures to ensure the vehicle is geo-fenced within the ODD.”

This allocates the responsibility of geofence verification to the LLP module, emphasizing multiple independent sensor modalities—GPS, camera, radar, and potentially LiDAR, and corresponding perception algorithms. The objective is continuous confirmation that the vehicle’s real-time position remains inside a validated operational domain.

\textbf{Step 2. Derivation from AI Safety Properties}
To refine this high-level intent, three core AI safety properties, robustness, reliability, and bias/fairness, are applied to the LLP ODD detector.

\begin{itemize}
    \item \textbf{Robustness:} Requires the model to maintain correct geofence classification despite input perturbations (e.g., GPS drift, camera noise, moderate weather). Accordingly, the LLP shall sustain less than 1\% degradation in boundary detection accuracy under GPS position errors of $\pm$5 m, blurred or noisy camera frames, and light–moderate rain or fog.
    
    \item \textbf{Reliability:} Demands consistent performance across time and nominal conditions. Thus, the module must achieve $\geq$99\% correct ``in-ODD vs. out-of-ODD'' classifications during day/night driving, with no more than one false classification per ten hours of continuous operation.
    
    \item \textbf{Bias \& Fairness:} Ensures equitable performance across all sub-regions of the ODD (urban, suburban, rural) and environmental contexts (dry vs. wet roads). The LLP’s classification accuracy may not deviate by more than $\pm$2\% between any of these scenarios, thereby preventing systematic degradation in any particular segment of the validated domain.
\end{itemize}

\textbf{Step 3. Derivation from Safety Analysis}
A system-level safety analysis (e.g., STPA or FTA) identifies hazards associated with misclassification of geofence boundaries. In particular, failing to detect an imminent exit from the ODD (false negative) could allow the ADS to continue in an unsupported environment, leading to unsafe operation. Based on this hazard, the following requirement is derived:\\
"The LLP shall compute an ‘inside-ODD confidence score at $\geq 5 Hz$ by fusing multi-modal sensor outputs. If the confidence falls below 0.80 at any instant, the ADS shall initiate a Safe State request—triggering a driver takeover alert and controlled deceleration—within 100 ms."

This ensures timely transitions to a safe fallback whenever uncertainty breaches a specified threshold.

\textbf{Step 4. Derivation from Functional Insufficiencies}
Functional insufficiencies arise when the AI system fails under specific operational conditions, such as sensor calibration drift or degraded input distributions. Two key requirements emerge:

\textbf{I.}	Periodic Sensor Calibration Self-Checks:
“Every 10 minutes, the LLP must assess sensor calibration. If the camera’s reprojection error exceeds 2 pixels or the GPS drift exceeds 10 m, the system shall perform automatic recalibration or switch to a redundant sensing mode (e.g., radar-only geofence verification) before resuming full autonomous operation.”\\
This guards against cumulative drift that could gradually cause the system to misjudge its position relative to the geofence.\

\textbf{II.}	Cumulative Drift Monitoring:
“The LLP ODD detector shall continuously compare streamed vehicle position estimates against the HD map boundary. If a cumulative drift of more than 3 m is detected over a 30-second interval, the module shall alert the driver and decelerate to $\leq 10 km/h$ until calibration is restored.”

By monitoring long-term deviations, the system can detect scenarios in which individual self-checks may not trigger (e.g., gradual sensor misalignment) and still ensure safe fallback.

\textbf{Step 5. Derivation from On/off-Board Measures}\

\textbf{On-Board Measures}: To guarantee robust, vehicle-level execution, on-board fusion and degraded-mode requirements specify how the ADS shall behave when individual sensors fail or degrade:

“The LLP ODD system shall fuse GPS, camera, and radar data at a rate of at least $10 Hz$. If any modality fails to provide valid data for more than $200~ ms$, the fusion weights shall be adjusted dynamically. Should the resulting fused confidence drop below 0.75 for over $ 100~ ms$, the ADS shall transition into a \textit{Degraded Safe Mode}, imposing a reduced maximum speed and alerting the driver to take control..”\

This ensures that any single sensor’s outage or corruption does not lead to full reliance on a degraded or unsafe perception output. Instead, the system automatically adapts and issues an operationally safe fallback.\

\textbf{ Off-Board Measures}:
Off-board measures focus on maintaining up-to-date geofence and mapping data via cloud synchronization:

“The LLP ODD detection module shall synchronize HD map/geofence data with a central server at least once every 24 hours. If no successful update is retrieved within 24 hours, the ADS shall not operate in full autonomous mode until either a new update is obtained or an offline geofence dataset—with at least 95 

This requirement guards against stale or missing map information that could invalidate the geofence, ensuring that the vehicle does not rely on an outdated or incomplete representation of its operational domain.

\textbf{Step 6. Consolidated Requirements List}\\
It should be noted that this example is prepared as a proof of concept to demonstrate how safety requirements can be used to derive AI safety requirements. Additional requirements may be applicable as more details are incorporated into the analysis. Summarizing the above derivations yields the following refined AI safety requirements for an LLP-based ODD detection module, as shown in Table. \ref{Table:Consolidated Requirements}.
\begin{table}[ht]
\centering
\caption{Consolidated and Refined Requirements List}
\scriptsize
\begin{tabular}{|p{0.5cm}|p{6cm}|}
\hline
\textbf{Req. No.} & \textbf{Requirement Description} \\
\hline
1 & The LLP ODD detection module shall fuse at least three independent sensing modalities—GPS, camera, radar—in real time to verify geofence compliance. \\
\hline
2 & The LLP model shall sustain $< 1\%$ accuracy degradation under GPS drift of $\pm 5\,\text{m}$, moderate camera noise, and light–moderate weather conditions. \\
\hline
3 & The LLP ODD detector shall achieve $\geq 99\%$ classification accuracy (in-ODD vs. out-of-ODD) under nominal day/night conditions, with no more than one false classification per ten hours of continuous operation. \\
\hline
4 & The LLP model shall exhibit $\leq 2\%$ accuracy deviation across all ODD sub-regions (urban, suburban, rural) and environmental contexts (dry, wet). \\
\hline
5 & The LLP shall compute an ``inside-ODD confidence score'' at $\geq 5\,\text{Hz}$. If confidence $< 0.80$, the ADS shall request a Safe State—triggering a driver alert and controlled deceleration—within 100 ms. \\
\hline
6 & Every 10 minutes, the LLP shall assess camera calibration (reprojection error $\leq 2$ pixels) and GPS accuracy (drift $\leq 10\,\text{m}$). If thresholds are exceeded, it shall invoke automatic recalibration or switch to redundant sensing before resuming full autonomy. \\
\hline
7 & The LLP shall continuously compare vehicle position estimates against the HD map boundary. Upon detecting $> 3\,\text{m}$ cumulative drift over 30 seconds, it shall alert the driver and decelerate to $\leq 10\,\text{km/h}$ until calibration is restored. \\
\hline
8 & The LLP shall fuse GPS, camera, and radar at $\geq 10\,\text{Hz}$. If any modality’s data gap exceeds 200 ms, fusion weights must adjust dynamically; if fused confidence $< 0.75$ for over 100 ms, the ADS transitions to ``Degraded Safe Mode'' (reduced speed, driver alert). \\
\hline
9 & The LLP ODD detector shall synchronize HD map/geofence data with the cloud at least every 24 hours. Absent a valid update, the ADS shall not operate in full autonomy until a new update is obtained or an offline dataset ($\geq 95\%$ overlap) verifies the geofence. \\
\hline
\end{tabular}
\label{Table:Consolidated Requirements}
\end{table}
\\
These requirements collectively ensure that the LLP-based ODD detection module satisfies the original SOTIF requirement while addressing AI-specific safety properties, mitigating identified hazards, compensating for functional insufficiencies, and incorporating both on-board and off-board safeguarding strategies. By adhering to this structured derivation, the ADS can continuously verify its operational domain, promptly detect and handle uncertainty or sensor failures, and maintain safe behavior even under adverse or evolving conditions.\\

  \subsubsection{Safety Analysis on AI System}  
The primary objectives of the safety analysis on the AI system are threefold.
First, it aims to detect safety-related faults and AI errors that could compromise AI safety requirements.
Second, it seeks to analyze the underlying causes of these faults and errors.
Finally, it supports the development of appropriate safety measures designed to prevent or mitigate such issues. These measures may include improvements to AI design, methodologies, dataset generation, and updates to both AI safety requirements and related development processes. Additionally, the output of safety analysis supports the verification of AI safety requirements by refining or establishing new requirements concerning data specifications, data collection, design parameters, and testing protocols. In the previous subsection, we have shown how to use the output of AI safety analysis to derive new AI safety requirements. Some recommended safety analysis methods that can be applied to AI model can be listed as: Fault Tree Analysis (FTA) \cite{ali2020analyzing}, Failure Mode and Effect Analysis \cite{pandya2025fmea}, STPA \cite{kou2025unistpa}, Event Tree Analysis (ETA) \cite{kramer2020identification}, Bayesian Network \cite{deng2025quantitative, talluri2025enhancing}, and HAZOP \cite{molloy2024hazard} .\\

 \subsubsection{Data Management}  
Data has a critical impact on the performance of AI systems. ISO/PAS 8800 proposes a dataset lifecycle for AI system development, covering key activities such as data requirements, design, implementation, safety analysis, verification, validation, management, and maintenance. Its goal is to identify dataset insufficiencies that could affect system safety, establish relevant data-related safety properties, and propose countermeasures through dataset safety analysis at various lifecycle stages. A critical emphasis is placed on establishing a clear, bidirectional link between each AI safety requirement and its corresponding dataset requirements, ensuring traceability and alignment between data quality and system safety. In \cite{abbaspour2025dataset}, we explained the safety aspects of the dataset management for the ADS application in detail. FuSa and SOTIF requirements can influence dataset requirements through AI safety requirements, which are used to derive specific dataset requirements. ISO/PAS 8800 also outlines the necessary data-related work products to support safety assurance and applies to AI systems utilizing supervised, semi-supervised, or unsupervised learning.\\

 \subsubsection{Verification \& Validation}  
 Verification \& Validation (V\&V) ensures that the AI system meets its defined safety requirements through both verification and validation processes. It verifies that the AI system fulfills its internal safety requirements and validates that these requirements are effectively achieved when the AI system is integrated into the broader system. During the AI system verification and validation (V\&V) phase, stand‑alone performance is assessed, and testing is conducted at both the system and component levels, including the AI model and its pre‑ and post‑processing elements. The term “AI system safety validation” is also defined for this context, distinguishing it from its broader usage in general safety engineering.\\

\subsubsection{Monitoring \& Change Management}  
Monitoring AI system during operation is required for maintaining AI safety after deployment through continuous monitoring and risk management. It establishes the need for defined processes to ensure ongoing safety assurance, leveraging the measures from safety assurance argument and potentially introducing additional ones to identify and manage safety risks during operation. Furthermore, it emphasizes the importance of having response mechanisms in place to address unacceptable safety risks and mandates re-approval of any modified AI system before it is released back into operation.\\

\section{Discussion}
Assigning AI-based perception modules a QM designation often fosters the misconception that these components pose no safety-critical hazards and therefore warrant no further scrutiny. In traditional SAE Level 2 systems, this assumption is reinforced by the expectation that a supervising driver will promptly intervene in the event of a system fault, thereby mitigating concerns related to controllability \cite{ISO26262}. However, HOD features undermine this safeguard, as drivers may be unaware of the specific ODD conditions required for safe activation. Our LLP example for HOD clearly illustrates that a false-positive ODD detection can erroneously activate HOD in unsupported contexts, resulting in a SOTIF hazard (where severity $S>0$ and controllability $C>0$) a risk that a QM-only assessment would likely overlook.

Such scenarios necessitate a comprehensive safety analysis, including top-down approaches (e.g., CTA), bottom-up methods (e.g., functional insufficiency effect analysis), and scenario-based validation, to uncover latent failure modes and develop targeted mitigation strategies. Consequently, any AI perception component, even if rated under QM, must undergo a full SOTIF assessment. If SOTIF-related hazards are identified, adherence to ISO/PAS 8800 guidelines becomes essential for AI components. ISO/PAS 8800 outlines a rigorous AI safety lifecycle, detailing how to derive AI safety requirements, conduct safety analyses of both the AI model and its dataset, and maintain traceability between AI safety requirements, dataset specifications, and measurable validation criteria, ensuring that no unacceptable risks infiltrate the vehicle’s operational environment.\\

To ensure these AI safety requirements satisfy SOTIF risk acceptance criteria, the loop is closed by quantitatively mapping each requirement to hazard rates and aggregated system risk. First, each refined requirement is allocated a measurable performance metric (e.g., boundary classification error rate, Safe State trigger latency). During validation, these metrics are statistically assessed across extensive ODD test scenarios to estimate the residual risk of geofence violation (e.g., collisions or unsafe exits). This empirical risk is then compared against the SOTIF target (e.g., $\leq 1 \times 10^{-6}$ undesired events per kilometer). If the observed risk exceeds the criterion, the AI safety requirements—such as confidence thresholds or fusion redundancies, are iteratively tightened (e.g., raising the confidence threshold from 0.80 to 0.85, increasing fusion modality update rates). This feedback loop, linking —requirement-derived metrics, validation results, and SOTIF targets, ensures the LLP-based ODD detector remains verifiably within the accepted risk envelope for all validated operational conditions.

\section{Conclusion}
As the integration of AI into safety-critical driving automation systems accelerates, traditional safety paradigms must evolve to address emerging risks. 
This paper has demonstrated that even when a component is designated as QM, if it is associated with non-zero severity and controllability levels (i.e., $S > 0$ or $C > 0$), it can pose significant safety risks. Through theoretical examination and real-world case studies, we have shown that AI-driven perception systems, often developed under QM assumptions, can introduce hazards that demand structured risk assessments. This emphasizes the necessity of applying the rigor of ISO 21448 and, where AI complexity is involved, the additional assurance frameworks provided by ISO/PAS 8800.\\
These findings underscore the urgent need to adopt a holistic, standards-aligned approach to safety analysis that encompasses all AI components, regardless of their initial classification. By aligning with emerging standards such as ISO/PAS 8800 and integrating AI-specific risk assessment methodologies, the automotive industry can better anticipate and mitigate unintended functional behaviors. Ultimately, revising safety frameworks to reflect the realities of AI integration is essential to ensuring robust, end-to-end safety assurance in modern automated driving systems.
\bibliographystyle{ieeetr} 
\bibliography{Bibliography}

\end{document}